\documentclass[prx,reprint,amsmath,amssymb,aps]{revtex4-2}
\usepackage{graphicx}
\usepackage{dcolumn}
\usepackage{bm}

\begin{document}

\preprint{APS/123-QED}

\title{Measurement of Many-Body Quantum Correlations in Superconducting Circuits}

\author{Kamal Sharma} 
\author{Wade DeGottardi}
\affiliation{Department of Physics and Astronomy, Texas Tech University, Lubbock, Texas 79416,USA}

\begin{abstract}
Recent advances in superconducting circuit technology have made the fabrication of large, customizable circuits routine. This has led to applications beyond quantum information including quantum simulation. A key challenge in this effort is the identification of the quantum states realized. Here, we propose a probe circuit capable of measuring many-body correlations in an analog quantum simulator superconducting circuit. This probe, which is specially designed for many-photon states, exploits the non-linearity of the Josephson junction to measure two-point correlation functions of the superconducting phase operator. We demonstrate the capabilities of this design in the context of an LC-ladder with a quantum impurity. The proposed probe allows for the measurement of inherently quantum correlations, such as two-mode squeezing, and has the potential to significantly expand the scope of analog quantum simulations using superconducting circuits.
\end{abstract}

\maketitle
\section{Introduction}
Superconducting circuits are macroscopic objects capable of exhibiting quantum behavior~\cite{leggett_quantum_1985, martinis_energy-level_1985,vool_introduction_2017} and are the leading contenders in the race to achieve large-scale quantum computation~\cite{blais_circuit_2021, weber_coherent_2017, tennant_demonstration_2022}. Dramatic advances in the technology associated with their fabrication, control, and measurement~\cite{devoret_superconducting_2013, kjaergaard_superconducting_2020} have made customized superconducting circuits widely available for a variety of applications outside of quantum computation such as single-photon detection~\cite{goltsman_picosecond_2001}, quantum metrology~\cite{xiao_precision_1987}, materials testing~\cite{LU19943361}, and quantum-enhanced gravitational wave detection~\cite{tse_quantum-enhanced_2019}. Of particular interest is their use as quantum simulators. Historically, discoveries in condensed matter have required access to new materials and measurement techniques. Superconducting circuits are poised to challenge this paradigm by allowing the creation of engineered quantum many-body systems~\cite{meiser_superstrong_2006, kuzmin_superstrong_2019}. 

Simulations of many-body systems using superconducting circuits fall into two broad classes: digital~\cite{smith_simulating_2019,burger_digital_2022,fauseweh_digital_2021} and analog~\cite{braumuller_analog_2017}. In digital simulations, the unitary dynamics of a target quantum system is calculated in discrete steps for a desired time interval. Digital simulations have been used to explore the one-dimensional Hubbard model~\cite{arute2020observation} and quantum walks~\cite{yan_strongly_2019}. In analog simulations, the focus of interest here, circuit architecture realizes a desired Hamiltonian~\cite{weiss_variational_2021}. Analog simulations have realized Mott insulating behavior~\cite{ma_dissipatively_2019}, many-body localization~\cite{mehta_down-conversion_2023}, the superconductor--insulator transition~\cite{kuzmin_quantum_2019}, quantum impurities~\cite{mehta_down-conversion_2023, kuzmin_inelastic_2021}, and hyperbolic lattices~\cite{kollar_hyperbolic_2019, grimsmo_squeezing_2017}. Additionally, there have been numerous proposals to simulate photon pair condensates~\cite{cian_photon_2019}, hard-core bosons~\cite{yanay_two-dimensional_2020}, Dicke chains~\cite{zhang_quantum_2014}, superfluids~\cite{zheng_superfluidmott-insulator_2017}, and supersolids~\cite{jin_photon_2013}.

\begin{figure}
\centering
 \includegraphics[width = 75mm]{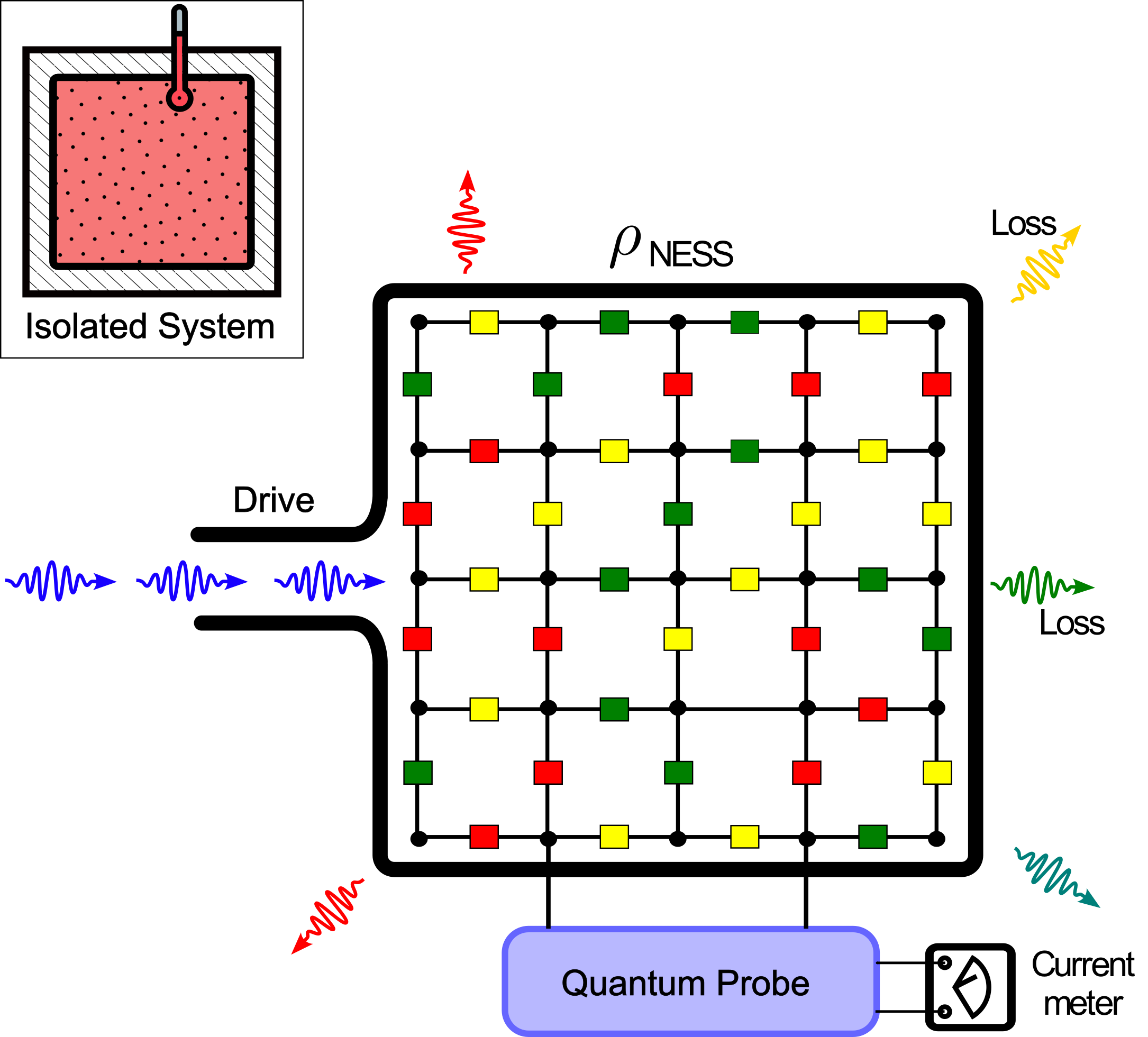}
 \caption{\textbf{A driven superconducting circuit.} The circuit is composed of nodes (black dots) connected by a capacitor, inductor, or Josephson junction (colored boxes). Superconducting circuits are open quantum systems driven far from equilibrium. The steady-state of the system takes a non-universal form that depends sensitively on the external drive, circuit parameters, and loss. In contrast, the measurement of condensed matter systems (inset) typically involves measuring the response of a weakly perturbed ground state or thermal state.}
 \label{fig:intro}
\end{figure}

A key challenge in analog quantum simulations is identifying the many-body state realized. The measurement of a traditional condensed matter system typically involves probing a system in its ground state or in a thermal state by weak perturbations. Through Linear response theory and the fluctuation-dissipation theorem allow the response of the system to be related to correlation functions of the relevant operator ~\cite{fetter_quantum_2003}. In fact, the behavior of a many-body correlation function is often taken to define a quantum phase of matter. For example, the existence of off-diagonal long-range order (ODLRO) formally defines the superfluid state~\cite{anderson_more_1972}.

However, unlike most condensed matter systems, superconducting circuits are lossy and typically probed far from equilibrium, as illustrated in Fig.~\ref{fig:intro}. Hence, their steady-state is neither the ground state nor a thermal state. Specially adapted measurement techniques are used to probe these non-equilibrium steady-states (NESS). For example, reflectivity measurements are used to measure the spectra of circuits~\cite{mehta_down-conversion_2023, kuzmin_inelastic_2021,ma_dissipatively_2019}. Other physical quantities measured experimentally include quadratures of the superconducting phase~\cite{hacohen-gourgy_continuous_2020, roy_quantum_2015} and photon populations~\cite{schuster_resolving_2007-1}. However, the ability to measure many-body correlations, which is crucial to the identification of quantum phases, has not been demonstrated in superconducting analog simulations. 

While classical quantities always have well-defined values, quantum degrees of freedom possess intrinsic uncertainties. For a generic operator $\hat{x}$, the quantity $\langle \hat x^2 \rangle$ is not equal to $\langle \hat x \rangle^2$, where $\langle .\rangle$ denotes the quantum mechanical expectation value. This also occurs for two correlated operators $\hat{x}$ and $\hat{y}$: $\langle \hat x \hat y \rangle$ differs from $\langle \hat x \rangle \langle \hat y \rangle$, in general. A consquence of this is that certain types of quantum order can only be detected through measurements involving the product of two operators. For example, this is the case for multi-mode squeezing~\cite{walls_quantum_1994}.

In this work, we propose a probe circuit capable of measuring correlation functions of two superconducting phase operators. This probe satisfies three general design criteria that we articulate. The capabilities of this probe are investigated in the context of a quantum impurity in a photonic cavity, a system that has attracted recent attention~\cite{mehta_down-conversion_2023, kuzmin_inelastic_2021}. In an inelastic collision with the impurity, a photon splits into several lower energy photons in a process known as photon down conversion. Photons created in this manner exhibit multi-mode squeezing~\cite{walls_quantum_1994} that can be detected by the proposed probe. Here, we illustrate the signatures of this effect with trial many-body wave functions.  

This paper is organized as follows. In Sec.~\ref{sec:design}, we establish the general criteria that a many-body superconducting probe must satisfy. A particular probe design, tailored for the cavity-impurity system, is introduced in Sec.~\ref{sec:photon_cavity}. In Sec.~\ref{sec:many_body_correlations}, trial wave functions exhibiting correlations characteristic of inelastic photon scattering are introduced and used to demonstrate the probe's capabilities. Finally, Sec.~\ref{sec:conclusion} summarizes our findings and discusses the future outlook.

\section{Probe Design}
\label{sec:design}

\begin{figure*}
\centering
\includegraphics[width=\linewidth]{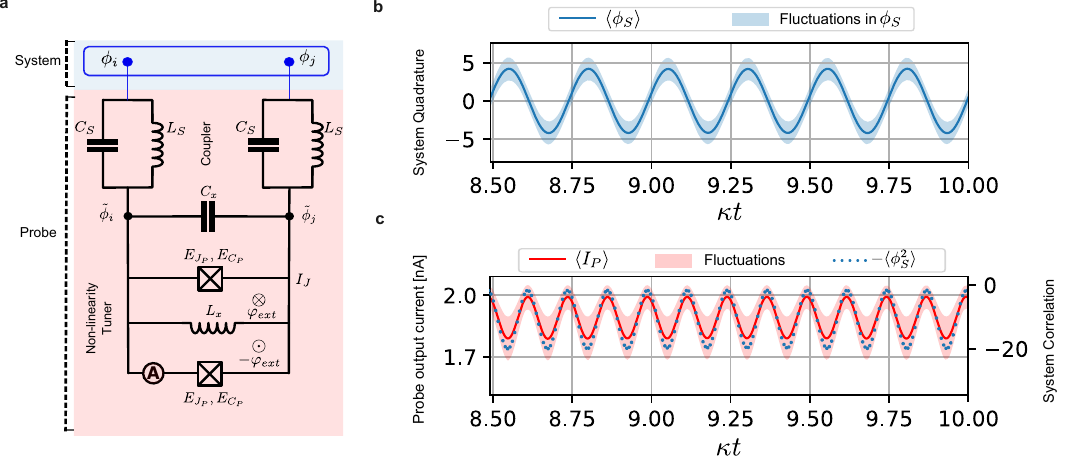}
\caption{\textbf{Probe circuit.} \textbf{a,}~Proposed probe circuit (red) connected to two nodes $i$ and $j$ of a generic superconducting circuit (blue). As described in the text, the output of the probe is controlled by an external magnetic flux $\varphi_{ext}$. The output signal of the probe is contained in the current through the lower Josephson junction and can be measured by an ammeter, as indicated by \textcircled{\raisebox{-.9pt} {\textbf{\sffamily A}}}.  \textbf{b,}~Calculated quadrature $\langle \phi_S \rangle$ as a function $\kappa t$, where $\kappa$ is the loss rate and $t$ is time, for a simulated system consisting of a single degree of freedom. In this example, the system contains approximately 7 photons. The shading indicates the uncertainty associated with quantum fluctuations. \textbf{c,} Output current (scale on left) for the simulation in \textbf{b}. For $\varphi_{ext}=\pi/2$, the output current (solid red, scale on left) is approximately proportional to $\langle\phi_S^2\rangle$ (dotted blue, scale on right). The magnitude of the quantum fluctuations in the probe's output (shaded red) is controlled by the average number of photons in the probe.}
\label{fig:probe_circuit}
\end{figure*}

A superconducting circuit is specified by a set of nodes that can be connected by one or more of three circuit elements: an inductor, a capacitor, or a Josephson junction (see Fig.~\ref{fig:intro}). The quantum dynamics of a circuit is captured by two canonically conjugate variables for each node. On the $i^{\textrm{th}}$ node, these variables are the superconducting flux $\Phi_i = \int_{-\infty}^t V_i (\tau) d \tau$ and its canonically conjugate charge $Q_i$. The quantity $V_i(\tau)$ is the time-dependent potential on this node. We introduce the dimensionless operators $\phi_i$, the superconducting phase of the node, and the number of Cooper pairs $n_i$, where $\Phi_i = \Phi_0 \phi_i / 2 \pi$ and $Q_i = 2 e n_i$. The constant $\Phi_0$ is the superconducting flux quantum and $e$ is the charge of the electron. For further discussion, see Appendix~\ref{appendix_system} and Ref.~\cite{girvin_113circuit_2014}. The goal of this paper is to introduce a probe circuit capable of measuring correlation functions of the superconducting phases $\langle \phi_i \phi_j \rangle$ in the steady state, where the expectation value $\langle \hat{X} \rangle = \mbox{Tr} ( \rho_{\textrm{NESS}} \hat{X})$ and $\rho_{\textrm{NESS}}$ is the non-equilibrium steady-state (NESS) density matrix. Such a probe will need to be customized to the system being measured. However, we identify three general criteria that such a probe should satisfy.

\begin{enumerate}
\item \textit{The probe should not alter the dynamics of the system.} Here, we consider probes that carry out weak continuous measurements of a system's steady state. In order for the probe not to fundamentally alter the system, it should perturb the system only weakly. For a probe that is coupled to the system in parallel, this condition is satisfied as long as the impedance of the probe $Z_P$ greatly exceeds the characteristic impedance of the system $Z_0$. For Hamiltonians that are periodic in the phase variables $\{ \phi_i \}$, as occurs for systems with Josephson junctions but no inductors~\cite{girvin_113circuit_2014}, care must be taken not to include any inductors in the probe that ``pin'' the phases of the system. Additionally, for systems that have ground-state degeneracies or spontaneously broken symmetries, the probe should not favor one ground state over the other.

\item \textit{The probe must contain a non-linear circuit element}. A probe that consists of linear circuit elements has currents and voltages that are linear in the circuit's degree of freedom, i.e., are functions of $\langle \phi_i \rangle$ and $\langle n_i \rangle$. Since $\langle \phi_i \phi_j \rangle \neq \langle \phi_i \rangle \langle \phi_j \rangle$ in general, such a probe cannot be used to obtain the desired correlation functions. 

\item \textit{The probe's output must be measurable.}  The probe's output must be measurable with current technology. If the probe's output is contained in a current, for example, then an output on the scale of nanoamps is measurable with current technology~\cite{lotkhov_dc_2019, grover_fast_2020}. A related consideration is the frequency response of the probe. For instance, if the probe's resonant frequency were to lie within the system's bandwidth, then the probe's response would be dominated by those modes resonant with the probe. To prevent this, the probe's resonant frequencies must lie outside the system's bandwidth. Additionally, fluctuations arising from noise and quantum effects must not obscure the signal. In order to limit quantum fluctuations associated with the probe, it is desirable for the probe to operate with as many photons as possible. This is accomplished by taking the probe's resonant frequencies to be as small as possible, and so they should lie \emph{below} the system's bandwidth.
\end{enumerate}

We propose a probe circuit, shown in Fig.~\ref{fig:probe_circuit}\textbf{a}, capable of satisfying these three criteria for the appropriate circuit parameters. The probe is connected to two nodes of the system, denoted $i$ and $j$. In accordance with criteria 2, the probe contains Josephson junctions. The key principle of the probe is that the correlation $\langle \phi_i \phi_j \rangle$ may be obtained from the current through the junctions with the appropriate flux biasing. The current through the two junctions is $I_{c} \sin{( \Delta \tilde{\phi}_{ij} \pm \varphi_{ext})}$, respectively, where $\Delta \tilde{\phi}_{ij} = \tilde{\phi}_i - \tilde{\phi}_j$, see Fig.~\ref{fig:probe_circuit}\textbf{a}. For $\varphi_{ext} = \pi/2$, the signal current $\propto \cos{(\Delta \tilde{\phi}_{ij})}$, which for $\Delta\tilde{\phi}_{ij} \ll 1$ gives a current $\propto 1 - \langle(\Delta \tilde{\phi}_{ij})^2 \rangle/2 $. This is a key intermediate result.

For $\varphi_{ext} = \pi/2$, the potential terms from the Josephson junctions cancel each other and thus do not enter into the equations of motion of $\tilde{\phi}_i$ or $\tilde{\phi}_j$, as shown in Appendix~\ref{appendix:probe}. The equations of motion give $\Delta \tilde{\phi}_{ij} = \beta \Delta{\phi}_{ij}$. According to criteria 3, the probe will be driven by frequencies far above its resonant frequency. In this limit, $\beta \approx 1/\left( 1 + 2 C_X/C_S \right)$. From the result in the previous paragraph, the linear relation $\Delta \tilde{\phi}_{ij} = \beta \Delta{\phi}_{ij}$ allows us to conclude that the current through either junction is $I_{c} = \left[ 1 - \beta^2 (\Delta\phi_{ij})^2 \rangle/2 \right]$, where $\Delta \phi_{ij} = \phi_i - \phi_j$. It is then straightforward to measure $\langle \phi_{i}^2 \rangle$ and $\langle \phi_j^2 \rangle$ by grounding one of the probe's terminals so that either $\phi_i$ or $\phi_j$ is equal to zero. The quantity $\langle \phi_{i} \phi_j \rangle$ is the cross term in $\langle \Delta \phi_{ij}^2 \rangle$, and thus can be obtained by subtracting $\langle \phi_{i}^2 \rangle$ and $\langle \phi_j^2 \rangle$ from $\langle \Delta \phi_{ij}^2 \rangle$.

The probe can also measure quadratures of the system such as $\langle \phi_i \rangle$. For $\varphi_{ext} = 0$, the current through either junction is $I_{c} \sin \Delta \tilde{\phi}_{ij}$. For $\Delta \tilde{\phi}_{ij} \ll 1$, the current is $\propto \Delta \tilde{\phi}_{ij}$, and thus can be used to measure $\langle \phi_{ij} \rangle$. Then, $\langle \phi_i \rangle$ and $\langle \phi_j \rangle$ are readily measured by grounding one of the probe's terminals. It should be noted that for $\varphi_{ext} = 0$, the junctions now enter into the equations of motion of the probe and must be accounted for. 

So far, it has been assumed that the dynamics of the system and probe is unitary. However, in large-scale superconducting circuits, photon loss plays a critical role. In order to assess the efficacy of the probe's design in the presence of loss, we consider a toy model in which the system is a single-mode resonator with a superconducting phase $\phi_S$. The dynamics of the system have been calculated using a Lindblad master equation and are shown in  Figs.~\ref{fig:probe_circuit}\textbf{b}, and \textbf{c}. (See Appendix~\ref{app:sim_loss} for details of the calculation.) The probe current closely tracks $\langle \phi_S^2 \rangle$ even in the presence of strong photon loss. As discussed in the context of criteria 3, the probe is subject to quantum mechanical fluctuations. These fluctuations have been calculated for the simulation and are shown in Fig.~\ref{fig:probe_circuit}\textbf{c}. For this simulation, the probe is in a coherent state and so the ratio of quantum fluctuations to the probe's output signal is set by the magnitude $\langle \Delta \tilde{\phi}_{ij}^2 \rangle$. When this quantity is larger, the relative size of the fluctuations are smaller. We expect that this feature will hold in general, i.e., even when the probe is not in a coherent state.

\section{Photon cavity with a quantum impurity}

\label{sec:photon_cavity}

\begin{figure*}
\centering
\includegraphics[width=\linewidth]{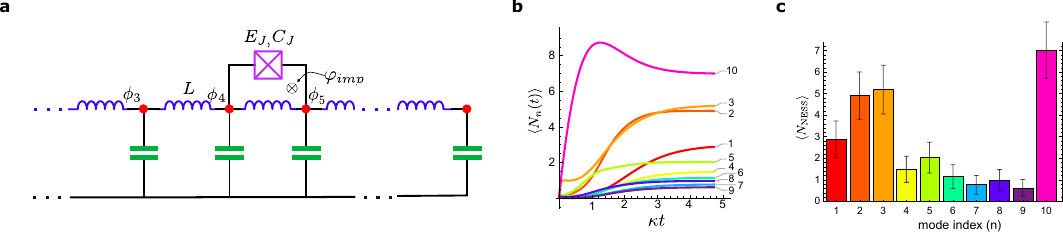}
\caption{\textbf{Dynamics of the cavity-impurity system.} \textbf{a,} Circuit diagram of an LC-ladder, which serves as a model for a photonic cavity, with a Josephson junction connected across nodes $i_0=4$ and $j_0=5$. The Josephson junction acts as a quantum impurity and induces inelastic photon scattering. \textbf{b,} Average photon number for the low frequency modes of the circuit. The tenth mode, which is excited by an external drive, overshoots its eventual steady-state value. \textbf{c,} Photon numbers in the steady state. The error bars indicate the 
magnitude of quantum fluctuations.}
\label{fig:sc-cavity}
\end{figure*}

In order to explore the capabilities of the proposed probe, we consider its ability to capture the physics of a simple system: quantum impurity in a photonic cavity \cite{mehta_down-conversion_2023, mehta_theory_2022}. Here, the cavity is modeled by an LC-ladder, which is shown in Fig.~\ref{fig:sc-cavity}\textbf{a}. When the number of nodes $N$ in the ladder is large, its low energy dispersion is linear and given by $\omega_n = n \omega_0$, where $n$ is a positive integer and $\omega_0 = \pi/((N+1) (LC)^{1/2})$. The Hamiltonian of the ladder is $H_0 = \sum_n \hbar\omega_n^{\phantom\dagger} a_n^\dagger a_n^{\phantom\dagger}$, where $a_n^{\phantom\dagger}$ ($a_n^{\dagger}$) is the photon annihilation (creation) operator for the $n^{th}$ mode (see Appendix~\ref{app_subsection:LC}). The superconducting phase on the $i^{th}$ site of the ladder can be written 
\begin{equation}
\phi_i = \sum_n \gamma_{in} \left( a_n^\dagger + a_n^{\phantom\dagger} \right),
\label{eq:mode_expansion}
\end{equation}
where $\gamma_{in}$ is the spatial wave function of the $n^{th}$ mode on the $i^{th}$ site.

A flux-biased Josephson junction, connected to sites $i_0$ and $j_0$ of the ladder, plays the role of the quantum impurity. The impurity Hamiltonian takes the form
\begin{equation}
H_{\textrm{imp}} = g \sum_{n m l} A_{nml} a_n^{\dagger} a_m^{\dagger} a_{l}^{\phantom\dagger} + h.c.,
\label{eq:anharmonic}
\end{equation}
where the coupling strength $g = E_J (E_C/E_L)^{3/4}$ and the parameters $A_{nml}$ are dimensionless constants of order unity. In a typical experimental setup, this system is excited with a coherent drive, as described by $H_{\textrm{drive}}$. The full Hamiltonian of the system is then $\mathcal{H} = H_0 + H_{\textrm{imp}} + H_{\textrm{drive}}$.

Initially, the system is in its ground state, with no photons in any of the modes. A drive, which is resonant with one of the modes of the ladder, is turned on. As the driven mode is populated, its photons scatter inelastically from the quantum impurity, giving rise to both up and down conversion. In down conversion, for example, a photon of frequency $10 \omega_0$ can decay into two photons of frequency $4 \omega_0$ and $6 \omega_0$. As various modes of the system become more populated, the rate at which photons are lost to the environment increases. Eventually, the system reaches a steady-state.

A simulation confirms this qualitative picture. The time-dependent photon population, shown in Fig.~\ref{fig:sc-cavity}\textbf{b}, was calculated using a master equation approach (see Appendix \ref{appendix_system}). This calculation captures the long-time dynamics of the system, but is not reliable at early times due to the breakdown of Fermi's golden rule, which is inapplicable for $t \ll 1/g$, where $g$ is the anharmonic coupling strength given in Eq.~(\ref{eq:anharmonic}). This explains the lack of coherent effects, such as Rabi oscillations, in Fig.~\ref{fig:sc-cavity}\textbf{b}. 

\section{Many-Body Correlations}

\label{sec:many_body_correlations}

\begin{figure*}
\centering
\includegraphics[width =  \linewidth]{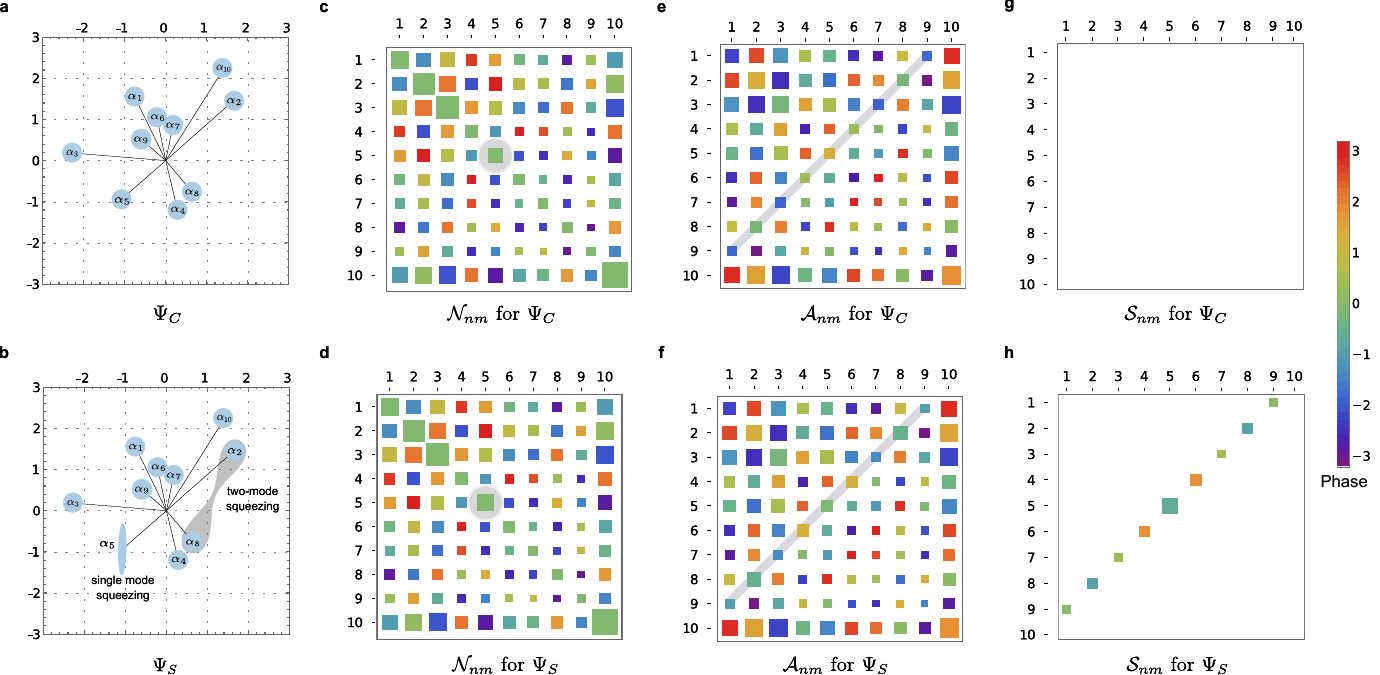}
\caption{\textbf{Many-body correlations.} \textbf{a,b}, Phasor diagrams for the the trial wave functions $\Psi_C$ and $\Psi_S$. The blue circles represent the single-mode coherent states with the corresponding complex parameters $\alpha_n$. The wave function $\Psi_S$ exhibits both two-mode and single-mode squeezing, as labelled. \textbf{c,} Hinton plot of the quantities $\mathcal{N}_{nm}$ for $\Psi_C$ and \textbf{d,} for $\Psi_S$. The area of each box indicates the magnitude of $\mathcal{N}_{nm}$ while its color indicates the phase delay in time (the color scale shows the phase in radians). For example, for $\Psi_C$, $\mathcal{N}_{5,5} = 2.0414$. \textbf{d,} Note that $\mathcal{N}_{5,5}=2.6052 $ for $\Psi_S$ is larger than $\Psi_C$ due to squeezing. \textbf{e,f,} Hinton plots for $\mathcal{A}_{nm}$ are shown for $\Psi_C$ and $\Psi_S$. The shaded region highlights the enhancement of the magnitude of $\mathcal{A}_{nm}$ due to two-mode squeezing. \textbf{g,} The quantity $\mathcal{S}_{nm}$ vanishes for $\Psi_C$ and in \textbf{f,} is non-zero along the super antidiagonal for $\Psi_S$.}
\label{fig:squeezed_state_measurements}
\end{figure*} 

Even for the modestly-sized system studied here, a full numerical calculation of the steady-state density matrix is impractical. We thus consider several trial density matrices that capture the essential physics. We first define $\rho_{\textrm{NESS}} = | \Psi_C \rangle \langle \Psi_C |$, where $| \Psi_C \rangle$ is an external direct product of coherent states, $|\Psi_C \rangle=\vert \alpha_1 \rangle \otimes \vert \alpha_2 \rangle\ldots\otimes \vert \alpha_{10} \rangle$, where $| \alpha_n \rangle$ is a coherent state for mode $n$ and $\alpha_n$ is the eigenvalue of the annihilation operator $a_n$ (see Appendix~\ref{appendix:trial_wavefunctions} for further details). This state, although classical, serves as an important benchmark to identify quantum effects. It also serves as a building block for non-trivial states. The complex numbers $\alpha_n$ are readily measured using the probe. As mentioned in Sec.~\ref{sec:design}, it is possible to measure quadratures $\langle \phi_i \rangle$ using the probe with its control flux $\varphi_{ext}$ set to zero. Using the mode expansion (\ref{eq:mode_expansion}), we see that $\langle \phi_i \rangle = \sum_n \gamma_{in} \mathcal{Q}_n(t)$, where $\mathcal{Q}_n(t) = \langle a_n^\dagger  + a_n^{\phantom\dagger}\rangle = \alpha_n e^{-i \omega_n t} + \alpha_n^\ast e^{i \omega_n t}$. Provided that the mode wave functions $\gamma_{in}$ are known, the magnitudes and phases of all the $\alpha_n$ can be obtained from a single measurement of $\langle \phi_i \rangle$ for all $n$ provided that $\gamma_{in}$ is non-zero. In Fig.~\ref{fig:squeezed_state_measurements}\textbf{a}, a phasor plot for a particular $| \Psi_C \rangle$ is shown. The magnitudes of the $\alpha_n$ have been selected to match the steady-state photon population found from the solution of the master equation while the phases of the $\alpha_n$ have been assigned at random. In an actual experiment, these phases contain important information about the system's approach to steady state.

A key signature of photon down conversion is squeezing. A multi-mode squeezed state $|\Psi_S\rangle$ is obtained by applying the two-mode squeezing operators $S(\xi_{nm})=\exp\left(\xi_{nm}^\ast a_n a_m - \xi_{nm} a_n^\dagger a_m^\dagger\right)$ to $| \Psi_C \rangle$ (see Appendix~\ref{appendix:trial_wavefunctions}). Squeezing arises in the cavity-impurity system because the propagator $\exp (-i H_{imp} \Delta t/\hbar)$ mimics the form of the squeezing operator $S(\xi_{nm})$. In fact, for the case that the driven mode ($n=10$) is a coherent state, $a_{10} \to \alpha_{10} e^{-i\omega_{10}t}$ and the propagator is given by $S(\xi_{nm})$ with the squeezing parameter $\xi_{nm} = i \alpha_{10} e^{-i\omega_{10}t} g A_{nm,10} t/\hbar$. For simplicity, we assume that the effects of the driven mode $n = 10$ dominate the squeezing and so the squeezing from other modes is not included in $| \Psi_S \rangle$. 

While quantities like quadratures $\langle \phi_i \rangle$ are unaffected by multi-mode squeezing, its effects can be detected through the measurement of correlation functions such as $\langle \phi_i^2 \rangle$ and $\langle \phi_i \phi_j \rangle$. In order to assess the effects of squeezing, it is helpful to express these correlation functions in terms of the expectation values
\begin{subequations}
\begin{eqnarray}
  \mathcal{N}_{nm}(t) &=& \langle a_n^\dagger a_m^{\phantom\dagger} + a_m^{\dagger} a_n^{\phantom\dagger} \rangle /2, \label{eqn:correlationsa} \\
  \mathcal{A}_{nm}(t) &=& \langle a_n^\dagger a_m^\dagger + a_m^{\phantom\dagger} a_n^{\phantom\dagger} \rangle/2. 
\label{eqn:correlationsb}
\end{eqnarray}
\label{eqn:correlations}%
\end{subequations}
For weak impurity scattering, the quantities $\mathcal{N}_{nm}(t)$ and $\mathcal{A}_{nm}(t)$ consist of single frequencies $(n-m)\omega_0$ and $(n+m)\omega_0$, respectively. Since $|\xi_{nm} | \propto t$, ostensibly the correlation functions associated with squeezing will grow without bound. In fact, an exact solution of a two-mode model shows that this does occur in a strong pumping regime~\cite{seifoory2016properties}. However, for weak impurity scattering, the squeezing parameter reaches a steady-state value. Based on the exact solution~\cite{seifoory2016properties}, we conclude that $\mathcal{N}$ and $\mathcal{A}$ are given by Eqs.~(\ref{eq:N}) and (\ref{eqn:squeezing}) with an effective $\xi_{nm} = i \alpha_{10}  e^{-i\omega_{10
}t} A_{nm,10} g T_\ast/\hbar$, where $T_\ast$ is a phenomenological parameter of the order of the time it takes the system to reach the steady-state. In Fig.~\ref{fig:squeezed_state_measurements}, the magnitude and phase of these quantities is plotted for the trial wave functions $| \Psi_C \rangle$ and $| \Psi_S \rangle$. The effects of two-mode squeezing are apparent if we compare $\mathcal{A}_{nm}$ for the $| \Psi_C \rangle $ and $| \Psi_S \rangle$, cf. Figs.~\ref{fig:squeezed_state_measurements}\textbf{e} and \ref{fig:squeezed_state_measurements}\textbf{f}). Additionally, the interaction (\ref{eq:anharmonic}) generates significant intra-mode squeezing for $n = 5$, which leads to an enhancement of $\mathcal{N}_{5,5}$ (cf. Figs.~\ref{fig:squeezed_state_measurements}\textbf{c},\textbf{d}). It is helpful to introduce
\begin{equation}
\mathcal{S}_{nm} = \mathcal{N}_{nm}(t) + \mathcal{A}_{nm}(t) - \mathcal{Q}_n(t) \mathcal{Q}_m(t)/2,
\end{equation}
a quantity that measures the strength of quantum correlations between various modes. This quantity vanishes for a product of coherent states, such as $\Psi_C$, see Fig.~\ref{fig:squeezed_state_measurements}\textbf{g}. On the other hand, it is non-zero for $\Psi_S$, as shown in Fig.~\ref{fig:squeezed_state_measurements}\textbf{h}.

An important measure of the probe's effectiveness is the number of distinct measurements of $\langle \phi_i \rangle$ or $\langle \phi_i \phi_j \rangle$ that are required to obtain all the quantities $\mathcal{Q}_n(t)$, $\mathcal{N}_{nm}(t)$, and $\mathcal{A}_{nm}(t)$. Provided that the photon wave functions $\gamma_{in}$ are known and non-zero, all $\mathcal{Q}_n$ can be obtained from $\langle \phi_i \rangle$ for a single site $i$, as discussed above. This assumes that the spectrum of $H_0$ is non-degenerate. In principle, the many-body correlations $\mathcal{N}_{nm}(t)$ and $\mathcal{A}_{nm}(t)$ could be obtained from $\langle \phi_i \phi_j \rangle$ for a single pair of sites $i$ and $j$. However, $H_0$ has a harmonic spectrum that gives rise to degeneracies. For example, both $\mathcal{A}_{3,4}$ and $\mathcal{A}_{5,2}$ have a frequency of $7 \omega_0$. In such cases, the required number of measurements will be equal to the largest degeneracy provided that each measurement contributes an equation in the various $\mathcal{N}_{nm}$ and $\mathcal{A}_{nm}$ that is linearly independent from the others. (The spatial symmetry of $\gamma_{in}$ can lead to redundant measurements). Then, the number of equations matches the number of unknowns and so all the correlation functions can be obtained. For the linear spectrum considered here, the largest degeneracy for either $\mathcal{N}_{nm}(t)$ and $\mathcal{A}_{nm}(t)$ goes as $\mathcal{O}(N)$, and thus $\mathcal{O}(N)$ independent measurements of $\langle \phi_i \phi_j \rangle$ are required. In this respect, the equally-spaced spectrum represents a worst-case scenario in terms of the number of measurements required due to the large number of degeneracies. For states with a small number of populated modes, such as the steady state considered here, only a few measurements are needed.

\section{Summary and Outlook}
\label{sec:conclusion}

We have proposed a probe circuit capable of measuring quantum correlations $\langle \phi_i \phi_j \rangle$ in analog quantum simulators built from superconducting circuits. The capabilities of the probe were explored for a cavity with a quantum impurity, a system that has garnered considerable recent attention~\cite{mehta_down-conversion_2023,kuzmin_inelastic_2021}. We focused on the probe's ability to detect the effects of squeezing arising from inelastic scattering from a quantum impurity. The detailed estimates of experimental parameters provided show that the probe is achievable with current technology.

While we have focused on a weakly coupled system, the proposed probe will be particularly important to the study of strongly coupled systems that may not be well understood theoretically. In such cases, the measurement of $\langle \phi_i \rangle$ would probe the system's quasiparticle spectrum, while measurement of correlation functions would potentially allow the identification of various types of quantum order. Recent proposals for analog quantum simulations of various physical systems are numerous, including the sine-Gordon model~\cite{roy_quantum_2021}, photon condensates~\cite{cian_photon_2019}, the Mott insulator to superfluid transition~\cite{zheng_superfluidmott-insulator_2017}, as well as supersolids~\cite{jin_photon_2013}. 
The ability to measure correlation functions has been a boon to quantum simulators based on cold atoms~\cite{gluza_quantum_2020,tajik_experimental_2023, tajik_verification_2023}. The proposed probe offers similar capabilities and would significantly expand the power and scope of superconducting circuit-based simulators. 

\begin{acknowledgments}
We thank Mohammad Hafezi and Brian Kennedy for their insightful comments.
\end{acknowledgments}

\appendix

\section{Probe Design}
\label{appendix:probe}

In this Appendix, we describe the equations of motion of the probe. The probe is connected to sites $i$ and $j$ of the system. The probe consists of two nodes, with superconducting phases $\tilde{\phi}_i$ and $\tilde{\phi}_j$, as shown in Fig.~\ref{fig:probe_circuit}\textbf{a} of the main text. In the limit that the impedance of the probe greatly exceeds that of the system, the phase difference in the system $\Delta \phi_{ij} \equiv \phi_i - \phi_j$ essentially acts as an external drive for the probe with limited back action.

The Josephson junctions contribute a term
\begin{equation}
\Delta \mathcal{L} = 2 E_{J,P} \cos \left( \varphi_{ext} \right) \cos \left( \Delta \tilde{\phi}_{ij} \right)
\end{equation}
to the Lagrangian. For $\varphi_{ext} = \pi/2$, this term vanishes and thus does not enter into the equations of motion, which are thus linear. The coupler stage of the circuit consists of a capacitor $C_S$ and inductor $L_S$ in parallel connecting nodes $i$ and $j$ of the system to those of the probe. The purpose of the coupler is to scale the incoming signal so that $\Delta \tilde{\phi}_{ij} \ll 1$, a necessity described in the text. From the Euler-Lagrange equations it is straightforward to show that $\Delta \tilde{\phi}_{ij} = \beta \Delta \phi_{ij}$ (see Fig.~\ref{fig:probe_circuit}). For a pure tone with $\Delta \phi_{ij} = A_n e^{i \omega t}$, we find that
\begin{equation}
\beta = \frac{1/L_S - C_S \omega^2}{ \left(1/L_S + 2/L_x \right) - \left( C_S + 2 C_x \right) \omega^2}.
\label{eq:beta}
\end{equation}
In the limit that $\omega$ greatly exceeds the resonance frequency of the probe, $\beta \approx 1/(1+ 2 C_X / C_S)$.

In Table~\ref{tab:probe1}, we provide estimates of probe parameters that are well within the reach of current experimental technology, see e.g.~\cite{mehta_down-conversion_2023}. These values, along with the parameters given for the system in Table~\ref{tab:system}, satisfy the three criteria given in the text.

\begin{table}
\begin{tabular}{@{}ll@{}}
    \toprule
    Probe Parameters $\qquad$& Value   \\
    \hline
    $C_S$       & $10$ pF \\
    $L_S$       & $1 \, \mu$H  \\
    $C_X$       & $10$ pF \\
    $L_X$       & $1 \, \mu$H \\
    $Z_P$       & $6$ k$\Omega$ \\
    $\omega_P$  & $100$ MHz \\
    $\beta$     & $0.3$     \\
    $I_{c,P}$   & $1$ nA \\
    \botrule
\end{tabular}
\caption{Circuit parameters of the probe designed to measure two-point correlations in the LC-ladder shown in Fig.~\ref{fig:sc-cavity}\textbf{a}}
\label{tab:probe1}
\end{table}

\section{Simulation of Probe With Loss}
\label{app:sim_loss}

The simulation of the probe, which was used to obtain Figs.~\ref{fig:probe_circuit}\textbf{b} and 2\textbf{c} assumes a simple model for the system that consists of a single driven mode. The Hamiltonian is 
\begin{eqnarray}
H/\hbar = \omega_{r} a_S^\dagger a_S^{\phantom\dagger} + \omega_{p} a_P^\dagger a_P^{\phantom\dagger} &-& i \Omega_0 (e^{i \omega_d t} a - h.c. ) \\ &+& \epsilon (a_P^\dagger a_S + a_S^
\dagger a_P). \nonumber
\end{eqnarray} 
where $\omega_S$ $(\omega_P)$ is the resonant frequency of the system (probe) and $\omega_d$ is the drive frequency. The frequency $\epsilon$ sets the size of the system-probe coupling and $\Omega_0$ is the drive strength. 

For a quantum system coupled to a bath, the time evolution of the density matrix $\rho$ is given by the Lindblad master equation~\cite{gardiner_quantum_2014}
\begin{equation}
\frac{\partial \rho}{\partial t} = - i \left[ \mathcal{H}, \rho \right] + \mathcal{L} \rho.
\label{eqn:evolution}
\end{equation}
At sufficiently low temperatures, the photons in the system are lost forever to the bath and do not scatter back into the system. 
Additionally, in the Markovian limit and for weak coupling to the environment, the Lindblad form of the master equation describes the photon loss from the system-probe  through the terms
\begin{equation}
    \mathcal{L} \rho = \sum_{i\in \{S,P\}}\kappa_i \left( {a_i} \rho {a}^\dagger_i - \frac{1}{2} \left\{ {a}^\dagger_i {a_i}, \rho \right\} \right),
\end{equation}
where $\{ A, B \} = AB + BA$. From the master equation, the evolution of the expectation value of a generic operator $G$ is~\cite{gardiner_quantum_2014} 
\begin{multline}
\frac{d\langle G(t)\rangle}{dt}=-\frac{i}{\hbar} \langle [G, H] \rangle+
\sum_{i\in \{S,P\}} \kappa_i \bigl(\langle a_i^\dagger G a_i\rangle\\ -\frac{1}{2} \langle\{a_i^\dagger a_i,G\} \rangle \bigr),
\label{eq:G}
\end{multline}
when the bath is in its ground state.

Figs.~\ref{fig:probe_circuit}\textbf{b,c} were obtained by solving the differential equation (\ref{eq:G}) for $G = a_S+a_S^\dagger$ and $G = I_c \cos(\phi_P)$, respectively. The parameters used are given in Table~\ref{tab:probe2}.

\begin{table}
\begin{tabular}{@{}ll@{}}
    \toprule
    Simulation parameters $\qquad$    & Value   \\
    \hline
    $\phi_{S}^{zpf}$         & 1  \\
    $\phi_{P}^{zpf}$         & 0.1   \\
    $\omega_S$     & $25$ GHz \\
    $\omega_d$     & $25$ GHz  \\
    $\Omega_0$     & $40$ GHz  \\
    $\omega_P$     & $12$ GHz  \\
    $I_{c,P}$     & $2.01$ nA \\
    $\epsilon$  & $10$ GHz \\
    $\kappa$    & $1$ GHz \\
    \botrule
\end{tabular}

\caption{Circuit parameters for single mode driven resonator and its probe.}
\label{tab:probe2}
\end{table}

\section{Cavity-Impurity System}
\label{appendix_system}

\subsection{Superconducting cavity}
\label{app_subsection:LC}

The photonic cavity is modeled by a ladder circuit of inductors and capacitors, as shown in Fig. \ref{fig:sc-cavity}\textbf{a} of the text. As is customary for superconducting circuits, we begin by writing down the Lagrangian of the circuit. For the ladder, we have 
\begin{equation}
    \mathcal{L}_{0}(\Phi_j,\dot{\Phi}_j) = \sum_{j=1}^{N} \frac{C \dot \Phi_j^2}{2} - \sum_{j=0}^{N}\frac{(\Phi_{j+1}-\Phi_{j})^2}{2L}.
\label{eqn:LC_lagrangian}
\end{equation}
To obtain the Hamiltonian, we use the fact that the conjugate momentum to $\Phi_j$ is $\partial \mathcal{L}_{0}/\partial \dot{\Phi}_j = C \dot{\Phi}_j$, which we identify as the charge $Q_j$ associated with the node. The Legendre transformation of the Lagrangian $\sum_j \Phi_j Q_j - \mathcal{L}_{0}$ results in the Hamiltonian $H_{0}$
\begin{equation}
H_{0}(\Phi_j,Q_j)= \sum_{j=0}^{N} \frac{Q_j^2}{2C} + \sum_{j=0}^{N}\frac{(\Phi_{j+1}-\Phi_{j})^2}{2L}
\label{eq:linear_LC_resonator}
\end{equation}
It is convenient to write the charge and flux operators as $Q_j = 2 e n_j$ and flux operators $\Phi_j = \frac{\Phi_0}{2\pi}\phi_j$. 
\begin{equation}
H_{0} = \sum_{j=1}^{N} 4E_C n_j^2 + \frac{1}{2}\sum_{j=0}^{N} E_L (\phi_{j+1}-\phi_{j})^2 ,
\label{eq:ham_app}
\end{equation}
where $E_C = \frac{4 e^2}{2C}$ and $E_L = \left(\frac{\Phi_0}{2\pi}\right)^2 \frac{1}{L}$. The dimensionless operators $n_j$ and $\phi_k$ obey the canonical commutation relations $\left[ n_j, \phi_k \right] =  i \delta_{jk}$, which follow from the definitions above.

We take the LC ladder to be terminated at its ends so that $\phi_0 = \phi_{N+1} = 0$. The classical solution of the flux obeying these boundary conditions is given by the normal modes and the allowed normal mode frequencies
\begin{eqnarray}
       X_{ni} &= \sqrt{\frac{2}{N+1}} \sin{\left(\frac{n\pi i}{N+1} \right) } , \label{eqn:coupled_HO_fixed} \\
       \omega_n &= \frac{2}{\sqrt{LC}} \sin{\left(\frac{n\pi}{2N+2}\right)} \label{eq:dispersion_LC_chain},
\end{eqnarray}
where, $X_n(j)$ is the spatial profile of the $n^{th}$ mode at node $j$. These eigenmodes satisfy the orthogonality relations
\begin{equation}\label{eq:orthonormality}
    \sum_{i} X_n(i) X_m(i) =\delta_{nm}.
\end{equation}
We introduce photon creation ${a}_n^{\dagger}$ and annihilation $({a}_n^{\phantom{\dagger}})$ in the mode expansion for $\phi_i$ in Eq.~(\ref{eq:mode_expansion}) and for
\begin{equation}
n_i = - i \left( \frac{E_L}{8 E_C} \right)^{1/2} \sum_n n \gamma_{in} \left( a_n^{\phantom\dagger} - a_n^\dagger \right),
\label{eq:n_expansion}
\end{equation}
where
\begin{equation}
\gamma_{in} = \frac{1}{\sqrt{n}} \phi_{zpf} X_{i n} 
\end{equation}
Substituting the expansions for $\phi_i$ and $n_i$ given by Eqs.~(\ref{eq:mode_expansion}) and (\ref{eq:n_expansion}), respectively, we find that the Hamiltonian~\ref{eq:ham_app} takes the form
\begin{equation}
H_{0} = \sum_n\hbar n\omega_0\left(a_n^\dagger a_n^{\phantom{\dagger}} + \frac{1}{2}\right).
\end{equation}
For large circuits with $N\gg1$, the low-lying modes are roughly equally spaced in frequency $\omega_n = n \omega_0,$ with
\begin{equation}\label{eqn:app_free_spectral_range}
    \omega_0 = \frac{\pi}{N+1}(LC)^{-1/2}.
\end{equation}

\subsection{Josephson junction impurity}

\label{sec:quantum_impurity}

The Hamiltonian of the Josephson junction impurity shown in Fig.~\ref{fig:sc-cavity}\textbf{a} is
\begin{equation}
H_{\textrm{imp}} =- E_{J} \cos{(\phi_{i_0} - \phi_{j_0} + \varphi_{\textrm{imp}})},
\end{equation}
where $E_{J,imp}$ is the Josephson energy and the junction is connected between the nodes $i_0$ and $j_0$ of the LC-ladder. The tunable magnetic flux $\varphi_{\textrm{imp}}$ changes the functional form of the non-linearity. When $\varphi_{\textrm{imp}} = \pi/2$, the leading terms in the expansion of $ H_{\textrm{imp}}$ gives the linear and cubic contributions
\begin{equation}
 H_{\textrm{imp}} \approx E_{J} \left[ (\phi_{i_0} - \phi_{j_0} ) - \frac{(\phi_{i_0} - \phi_{j_0} )^3}{3!} \right].
\label{eq:three-wave-mixing}
\end{equation}
Any odd power, in $\phi$, perturbation to the harmonic Hamiltonian vanishes to first order in perturbation theory. However, the cubic term in equation~\ref{eq:three-wave-mixing} induces the down conversion of photons even in the first order. In terms of the creation and annihilation operators, the impurity Hamiltonian is given by
\begin{equation}
H_{\textrm{imp}}=g \sum_{nml}^N A_{nml} (a_n^\dagger a_m^\dagger a_l + h.c.)
\end{equation}
where the energy scale of the coupling, $$ g = E_J\left(\frac{E_C}{E_L}\right)^{3/4},$$ appearing in Eq.~(\ref{eq:anharmonic}) is determined by the energies $E_J, E_C$, and $E_L$. The dimensionless coupling matrix  $A_{nml} \equiv A_{nml}(i_0,j_0) $ explicitly expressed as 
\begin{equation}
A_{nml}(i_0,j_0) = \left( 
 \sqrt{2/\pi}\right)^3 2^3 2^{3/4
} f_n f_m f_l,
\label{eq:Anml}
\end{equation}
where,
$$f_n = \frac{1}{\sqrt{n}} \sin{\frac{n \pi (i_0-j_0)}{2N+1}} \cos{\frac{n \pi (i_0+j_0)}{2N+1}}$$
depends on the set of nodes $\{i_0,j_0\}$ connected to the Josephson junction and determines the strength of coupling among the modes $n,m,l$. The rotating-wave approximation has been applied in deriving $H_{\textrm{imp}}$ and so only energy conserving terms satisfying $n\omega_0+m\omega_0=l\omega_0$ appear in the Hamiltonian~(\ref{eq:anharmonic}).

\subsection{Mode population dynamics}
\label{sec:app_time_evolution}

The time evolution of the photon population $\langle N_n \rangle$ for each mode $n$ can be calculated by solving the master equation for the diagonal elements of ${\rho}$. For any given mode $n$, the photon down-conversion process leads to a cascade of photons filling up and leaving the single-particle state corresponding to the mode $n$. For weak coupling between the modes, the rate of this process is determined by Fermi's golden rule 
\begin{multline}
\Gamma_{down} = \Gamma_0 \sum_{ml} A_{nml}^2 [(N_n+1) (N_m+1) N_l \\- N_n N_m (N_l+1) ] \delta({\Delta \omega}) 
\end{multline}
where $\Gamma_0 \sim E_J^2/(\hbar^2 \omega_0)$ and $\Delta \omega = \omega_0(n+m-l)$. down conversion continues from the newly populated states to more low-energy pairs, resulting in four photons. This continues till a steady-state is reached. 

The inevitable coupling of our system with the environment causes the photons to be lost from the circuit at a rate, $\kappa$. To account for photon loss, we work in the Markovian approximation. The evolution of the photon numbers can be derived from Eq.~(\ref{eq:G}) with $G(t) = a^\dagger a$. In this case, the action of the Liouville operator on the density matrix $\rho$ in Eq.~(\ref{eqn:evolution}) simplifies to
\begin{equation}
\label{eqn:eqn_Liouville}
\mathcal{L}\rho = - \kappa \sum_m N_m \vert N_m \rangle \langle N_m \vert.
\end{equation}
Finally the effect of the coherent resonant drive is to populate the mode of corresponding frequency. These three contributions are used to solve for the $N_{n}$ as a function of time. The master equation, thus, leads to a set of coupled differential equations for the mode population which can be solved numerically.
\begin{equation}\label{eq:master_equation}
\dot{N}_n = \Gamma_{down}  + \Omega_{\textrm{drive}} - \kappa N_n,
\end{equation}
where $\Omega_{\textrm{drive}} = \Omega_0 e^{-i \omega_{10}t} + \Omega_0^\ast e^{i \omega_{10}t}$. Here, the effective strength $\Omega_0$ of the driving tone is a function of the drive parameters and its coupling with the LC-ladder.

\subsection{Possible System Parameters}

In Table~\ref{tab:system}, we propose realistic experimental parameters for the cavity-impurity system. 

\begin{table}
\begin{tabular}{@{}ll@{}}
    \toprule
    System Parameters $\qquad$    & Value   \\
    \hline
    $N$         & 50   \\
    $L_{0}$     & $10.54$ nH \\
    $C_{0}$     & $0.58$ pF  \\
    $\omega_0$  & $300$ MHz \\
    $E_J/h$     & $94.5$ MHz \\
    $g/h$       & $5$ MHz \\
    $Z_0$       & $500 \, \Omega$ \\
    $\Omega_0$  & $18.12$ kHz \\
    $\kappa$    & $1.2$ kHz \\
    \botrule
\end{tabular}
\caption{Circuit parameters for the LC-ladder with weak impurity.}
\label{tab:system}
\end{table}

\section{Trial wavefunctions}
\label{appendix:trial_wavefunctions}
Here, we consider three trial wavefunctions that exemplify some of the many-body physics exhibited by large superconducting circuits. 

The wave function $\Psi_C$ is a multi-mode coherent state given by
\begin{equation}
\vert \Psi_C \rangle =\prod_{n=1}^{N} D_n(\alpha_n) \vert 0 \rangle
\end{equation}
 where
\begin{equation}
 D(\alpha) = e^{\alpha a^\dagger - \alpha^* a},
\label{eqn:displacement_operator}
\end{equation} 
is the displacement operator. The phase quadrature of $n^{th}$ mode is 
\begin{equation}
\mathcal{Q}_n (t) = \left( \alpha_n e^{-i \omega_n t}  + \alpha_n^\ast e^{i \omega_n t} \right).
\label{eq:Qn}
\end{equation}
The correlation functions defined in Eqs.~(\ref{eqn:correlationsa}) and (\ref{eqn:correlationsb}) are
\begin{eqnarray}
\mathcal{N}_{nm}(t) &=& \frac{1}{2} \left( \alpha_n^* \alpha_m e^{-i (\omega_m -\omega_n)t} + c.c.\right), \\
\mathcal{A}_{nm}(t) &=& \frac{1}{2} \left( \alpha_n^* \alpha_m^* e^{i (\omega_m +\omega_n)t} + c.c.\right) \label{eq:na},
\end{eqnarray}
respectively. Because these are coherent states, the operators $a_n$ and $a_n^\dagger$ behave like c-numbers and so the clustering holds for any correlation function. As a result, $\mathcal{S}_{nm}$ vanishes for $\Psi_C$.

The wave function $\Psi_S$ is characterized by two-mode squeezing and single-mode squeezing
\begin{equation}
\vert \Psi_S \rangle = \prod_{l} D_l (\alpha_n) \prod_{nm} S_{nm}(\xi_{nm}) | 0 \rangle.
\label{eq:psiM}
\end{equation}
where
\begin{equation}
S(\xi_{nm})=\exp\left(\xi_{nm}^\ast a_n a_m - \xi_{nm} a_n^\dagger a_m^\dagger\right)
\end{equation}
is the squeezing operator. Multi-mode squeezing does not affect $\mathcal{Q}_n$, which is given by Eq.~(\ref{eq:Qn}). However, $\mathcal{Q}_n$ is affected by single-mode squeezing with
\begin{equation}
\mathcal{Q}_n (t) = \left( \cosh | \xi_{nm}| \alpha_n e^{-i \omega_n t}  + \sinh|\xi_{nm} | \alpha_n^\ast e^{i \omega_n t} \right) \delta_{nm}.
\end{equation}

The two-operator correlation functions are
\begin{align}
\begin{split}
\mathcal{N}_{nm}(t)&=\frac{1}{2} \left( \alpha_n^* \alpha_m e^{-i \left(\omega_m -\omega_n \right)t} + \alpha_m^* \alpha_n e^{i (\omega_m -\omega_n)t} \right)\\ &+  \sinh^2\left( \vert \xi_{nm} \vert \right) \delta_{nm},
\label{eq:N}
\end{split}\\
\begin{split}\mathcal{A}_{nm}(t)&=\frac{1}{2} \left( \alpha_n^* \alpha_m^* e^{i (\omega_m +\omega_n)t} + \alpha_m \alpha_n e^{-i (\omega_m +\omega_n)t} \right) \\&- \cos \left( \arg \xi_{nm} \right) \sinh \left( \vert 2 \xi_{nm} \vert \right),
\end{split}
\label{eqn:squeezing}
\end{align}
These expressions assume that there is no higher order squeezing, i.e., no mode is squeezed with more than one other mode. 

For $\Psi_S$ in the paper, we have assumed that the effects of squeezing are dominated by the driven mode $n = 10$, while squeezing from the other modes has been neglected. To this end, we have applied two-mode squeezing to the pairs of modes $(1,9)$, $(2,8)$, $(3,7)$, $(4,6)$. Single-mode squeezing has been applied to mode $n = 5$. 
The equations for $\mathcal{N}$ and $\mathcal{A}$ given above show that single-mode squeezing affects $\mathcal{N}_{5,5}$, while two-mode squeezing alters the values of $\mathcal{A}_{1,9},\mathcal{A}_{2,8}$, and $\ldots\mathcal{A}_{9,1}$.

\providecommand{\noopsort}[1]{}
%


\end{document}